\documentclass[epj]{svjour}
\usepackage{graphicx,dsfont}
\begin{document}
\title{Excited hadrons on the lattice}
\subtitle{State of the art and future challenges}
\author{Christof Gattringer
\thanks{Invited plenary talk at NSTAR 2007, September 5-8 2007,
Bonn, Germany}%
}                     
%
%
\institute{Universit\"at Graz,
Institut f\"ur Physik,
FB Theoretische Physik,
 8010 Graz, Austria}
\date{Received: date / Revised version: date}
%
\abstract{
We review the techniques of lattice QCD calculations
for excited hadrons with light quarks and outline the future challenges that are faced in
calculations with fully dynamical fermions.  
} 
\maketitle
\section{Introduction}
\label{intro}
In the last 20 years the lattice approach has evolved into a powerful 
quantitative tool for obtaining non-perturbative results from 
ab-initio QCD calculations. A
revolutionary breakthrough was the understanding of chiral symmetry on the
lattice, which may be implemented by using a lattice Dirac operator that
obeys the Ginsparg-Wilson equation \cite{giwi}. Thus lattice QCD is now
conceptually ready for calculations with fully dynamical light quarks, and as
algorithms and computer technology are evolving rapidly, more challenging
problems are considered.

An prominent example for this type of more advanced calculations is the
analysis of excitations of hadrons. 
Excited states are much harder since they,
as we will outline in the next section, appear only in sub-leading terms of
the Euclidean correlators one analyzes on the lattice. Although powerful
methods are available for tackling this problem, it is still a challenge to
construct hadron interpolators that have a strong overlap with the
excitations, and at the same time may be implemented numerically in a 
cost-efficient way. 

About 5 years ago several groups started to systematically
explore hadronic excitations using lattice QCD \cite{earlyexcite}. 
A strong motivation for this
enterprise is, e.g., the unresolved nature of the first positive parity
excitation of the nucleon, the N(1440) Roper state. Although calculations at
relatively large volumes with light quarks were performed, there is still no
definitive answer to the Roper puzzle from the lattice. In particular the role
of dynamical sea quarks still has to be understood more clearly. Nevertheless,
the Roper problem has considerably pushed the development of the techniques 
for excited hadrons.

After a brief introduction to lattice methods, in this contribution
we review the recent developments and outline how a state of the 
art calculation
for excited hadrons with light quarks 
proceeds. We illustrate the current status by discussing a
few selected results in more detail.  

The above mentioned progress towards fully dynamical simulations with light
quarks gives rise to new challenges: So far most of the calculations were done
in the quenched approximation where excited states cannot decay. The same is 
true for a dynamical calculation when the quarks are heavy such that decays
are not possible due to kinematical reasons. However, slowly data from
simulations with light dynamical quarks become available and the problem of
particle decay and 
scattering states has to be faced. In the continuum a scattering state has a
continuous spectrum of energies and for a particular value of the relative
momentum its energy may be degenerate with the
energy of a bound state. In the Euclidean correlator one studies on the
lattice, this degeneracy leads to a mixing of scattering and 
bound states. The tools for disentangling the two types of states are ready in
principle - one uses a finite volume analysis - but the implementation of
these ideas in full lattice QCD calculations is still in its infancy. We
briefly address these problems, together with other future challenges, such as
the evaluation of matrix elements for excited states.

\section{Excited states in lattice QCD}

\subsection{Euclidean correlators on the lattice}

Before we come to discussing excited states on the lattice, we need to review
the central tool in lattice QCD, Euclidean correlators. In any lattice
calculation one evaluates numerically the correlators of some operators 
$\widehat{O}$. The Euclidean correlator of two operators $\widehat{O}_1$,
$\widehat{O}_2$ is defined as
\begin{eqnarray}
\langle O_2(t) \, O_1^\dagger(0) \rangle & = & 
\lim_{T\rightarrow \infty} \, \frac{1}{Z_T}  \mbox{Tr} \,
\big[ \widehat{O}_2 \, e^{-t \widehat{H}} \, \widehat{O}_1^\dagger \,
 e^{-(T-t) \widehat{H}} \big] , \; \; \; 
\label{ecorr} \\
Z_T & = & 
\mbox{Tr} 
\big[ e^{-T \widehat{H}} \big] \; .
\label{partition1}
\end{eqnarray}
In this expression $\widehat{H}$ is the QCD Hamiltonian and $t$ and $T$ are
real numbers, often referred to as Euclidean time. 

The interpretation of the Euclidean correlators is obtained with the help of 
the (unknown) 
eigenstates $| n \rangle$ of the Hamiltonian which obey
$\widehat{H} | n \rangle = E_n |n \rangle$, where $E_n$ denotes the energy 
of the eigenstate $|n\rangle$. If one uses them to compute the traces 
in (\ref{ecorr}) and (\ref{partition1}), and inserts the unit in the form 
$\mathds{1} = \sum_n |n \rangle \langle n |$ to the left of $\widehat{O}_1$
in (\ref{ecorr}), one finds
\begin{equation}
\langle O_2(t) \, O_1^\dagger(0) \rangle \; = \; 
\sum_n \langle 0 | \widehat{O}_2 | n \rangle 
\langle n | \widehat{O}_1^\dagger | 0 \rangle \, e^{-t E_n} \; .
\label{specsum}
\end{equation}
In this spectral representation for the Euclidean 2-point function, the sum
runs over all eigenstates $|n\rangle$ of the QCD Hamiltonian. Each term 
comes with a Boltzmann factor containing the corresponding energy $E_n$.
The energies are normalized relative to the vacuum state $|0\rangle$ with
energy  
$E_0 = 0$. The Boltzmann factors are multiplied with matrix elements of the
operators $\widehat{O}_i$ between the vacuum state $|0\rangle$ and a physical
state $|n\rangle$.

The spectral representation (\ref{specsum}) is a powerful tool. If one, e.g.,
wants to compute the mass of the nucleon, one uses for both $\widehat{O}_1$
and $\widehat{O}_2$ an operator $\widehat{O}_N$ with the quantum numbers of the
nucleon. For this choice the matrix element 
$\langle n | \widehat{O}_N^\dagger | 0 \rangle$ is non-vanishing only for
those terms in the sum, where $|n\rangle$ is the state $|N \rangle$
of the nucleon or one of its excitations $|N^\prime\rangle,
|N^{\prime\prime}\rangle\, ...\,$. Thus we obtain:
\begin{equation}
\langle O_N(t) \, O_N^\dagger(0) \rangle \; = \; 
\langle 0 | \widehat{O}_N | N \rangle 
\langle N | \widehat{O}_N^\dagger | 0 \rangle \, e^{-t E_N} \; + \; ... \; .
\label{twopointnucleon}
\end{equation}
Obviously we can extract the energy of the nucleon from the 
exponential decay of
the Euclidean correlator. Furthermore, the overall factor is related to the
matrix element of our operator with the physical nucleon state $|N\rangle$.

Having convinced ourselves, that the Euclidean correlator can be used to
extract energies and physical matrix elements, we need to discuss the way the
Euclidean correlators are evaluated in lattice QCD. Instead of the unknown
eigenstates of the QCD Hamiltonian, one uses eigenstates of the field
operators for computing the trace in (\ref{ecorr}). The exponential function
of the Hamiltonian is treated with the help of the Trotter formula, which
introduces an auxiliary direction that in the end gives rise to a
4-dimensional lattice formulation for the Euclidean correlation function,
given by
\begin{eqnarray}
&& \langle O_2(t) \, O_1^\dagger(0) \rangle \; = 
\label{ecorrlat} \\
&& \qquad = \; \frac{1}{Z} \int \!\!{\cal D}[G,\overline{q},q] \, 
e^{-S[G,\overline{q},q]} \,
O_2[G,\overline{q},q]_t \,  O_1^\dagger[G,\overline{q},q]_0 \; ,
\nonumber
\\
&& \;\; \; Z \; = \; 
\int \!\!{\cal D}[G,\overline{q},q] \, e^{-S[G,\overline{q},q]} \; .
\label{partitionlat}
\end{eqnarray}
The Euclidean correlation function is represented as a path integral over all
degrees of freedom, the gluons $G$ and the quark fields $\overline{q},\, q$.
All fields live on a 4-dimensional lattice, and the path integration 
${\cal D}[G,\overline{q},q]$ is implemented as the product of the integrals
over the classical degrees of freedom on all lattice points. 
Each configuration of the fields $G,\overline{q},q$ is
weighted with the Boltzmann factor $\exp(-S)$, where $S[G,\overline{q},q]$ is
a lattice discretization of the QCD action (derivatives are replaced by finite
differences on the lattice etc). In the path integral the operators 
$\widehat{O}_i$ appear as monomials $O_i$ of the classical field variables 
$G,\overline{q},q$. We will refer to these monomials as {\it interpolators}
from now on. The Euclidean time arguments $t$ and $(t=)\, 0$ 
determine from which time-slice of the 4-dimensional lattice the field
variables are taken for building the interpolators.  

The expression (\ref{ecorrlat}) has the form of an expectation value for a
statistical system in the canonical ensemble. Thus the numerical methods from
statistical mechanics, in particular Monte Carlo simulation techniques, 
can be taken over to lattice QCD. Using these we can evaluate the Euclidean
correlators numerically and extract the results for energies and matrix
elements using the spectral representation (\ref{specsum}). However, we
stress at this point, that the Monte-Carlo methods which are used for
evaluating the Euclidean correlators give rise to statistical errors. As we
will see below, this limited accuracy implies that special methods for
extracting the physical observables from the correlators have to be developed.

\subsection{Why are excited states so difficult?}

Having outlined the basic steps of a lattice calculation, we can return to the
problem of excited states. We have discussed that by selecting suitable
interpolators with the quantum numbers $I,J,P,C$ we can project onto the
states we want to analyze, and the example of  Eq.~(\ref{twopointnucleon})
illustrates how to study the positive parity nucleon channel. 
However, when we want to
consider the first positive parity excitation of the nucleon, the N(1440)
Roper state, we find that this state has the same quantum numbers
$\frac{1}{2}^+$ as the nucleon ground state. Thus all we
can obtain from our Euclidean correlator  
$\langle O_N(t) O_N^\dagger(0) \rangle$ is a sum of
exponentials 
\begin{eqnarray}
\langle O_N(t) \, O_N^\dagger(0) \rangle & \; = \; & 
\langle 0 | \widehat{O}_N | N \rangle 
\langle N | \widehat{O}_N^\dagger | 0 \rangle \, e^{-t M_N} 
\label{twopointnucleon2} \\
& \; + \; &
\langle 0 | \widehat{O}_N | N^\prime \rangle 
\langle N^\prime | \widehat{O}_N^\dagger | 0 \rangle \, e^{-t M_{N^\prime}} 
 ... \; \; ,
\nonumber
\end{eqnarray}
where $|N^\prime\rangle$ denotes the first excited state 
(with positive parity)  
and $M_{N^\prime}$ is the corresponding mass (from now on we use interpolators
projected to vanishing momentum, such that the energy reduces to the mass,
$E_i = M_i$). The dots indicate the contributions of higher excitations with
positive parity, such as the N(1710). We stress that negative parity states,
e.g., N(1535) and N(1650), do not contribute in the tower of excitations,
since the parity of our lattice interpolator $O_N$ was chosen positive. 

Eq.~(\ref{twopointnucleon2}) illustrates clearly why excitations are much
harder to extract: The spectral decomposition of the Euclidean 2-point
function is dominated by the exponential decay from the lightest mass
$M_N$. Only the sub-leading term contains the mass $M_{N^\prime}$
of the first excitation we  
want to study and this contribution is suppressed exponentially for
increasing $t$ with a factor of $\exp(-(M_{N^\prime} - M_N) \, t)$. This
implies that only for small values of $t$ we can hope to see a signal of the
first excitation. For higher excitations the situation is even worse. 

Although the signals of the excited states are suppressed exponentially,
several different techniques for extracting their spectrum have been applied
to the Euclidean correlators. In the next section we will discuss in detail
the variational method \cite{variation1,variation2}. However, 
also other approaches have been tried \cite{otherextract}. These 
attempts employ advanced fitting techniques or other analysis
tools for getting both the leading and the sub-leading
exponential or even try to reconstruct the spectral density
of the two point function. The results typically
suffer from the statistical errors which make multi-exponential fits 
rather unstable and from the fact that excitations produce significant
contributions only for small $t$. Consequently only a limited amount of 
information is available. In general the impression is that the alternative
methods \cite{otherextract} have not reached the quality of results from 
the variational method which we discuss in detail in the next section.

\section{Lattice technology for excited states}

\subsection{The variational method}

In the last section we have addressed the problem of extracting excited state
masses from the sub-leading exponentials of Euclidean correlators. The
decisive advantage of the variational method \cite{variation1,variation2}
is, that it extracts more
information from the system by analyzing more than a single correlator. 
Again we use our example of the nucleon to illustrate this
point. It is easy to check that the following 
two interpolators have the quantum
numbers of the nucleon ($C$ denotes the charge conjugation matrix),
\begin{eqnarray}
O_{N_1}(x) & \; = \; &
\epsilon_{abc} \left[u_a^T(x) \, C \gamma_5 \, d_b(x)\right ] u_c(x) \; ,
\label{chi1}
\\
O_{N_2}(x) & \; = \; &
\epsilon_{abc}\left[u_a^T(x) \, C \, d_b(x)\right ] \gamma_5 u_c(x) \; .
\label{chi2}
\end{eqnarray}
Consequently both of them should give rise to Euclidean correlators that can
be used to compute properties of the nucleon and of its excitations. 

\begin{figure}[t]
\begin{center}
\includegraphics[width=80mm,clip]{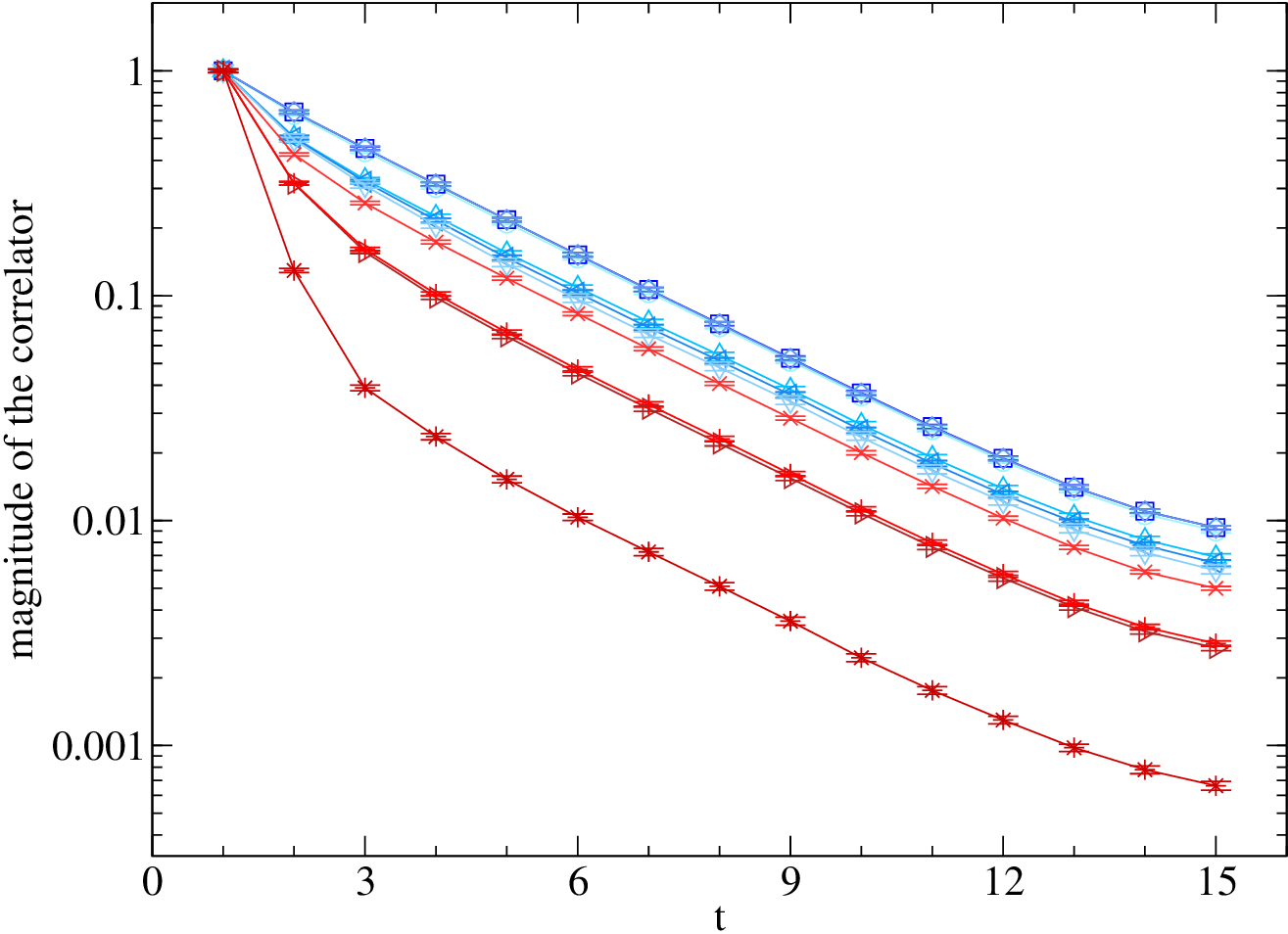}
\end{center}
\caption{Euclidean correlators for different interpolators with the quantum
  numbers of the pion. We plot the correlators as a function of Euclidean
  time $t$, using a logarithmic scale for the vertical axis.}
\label{fig1}       
\end{figure}

We illustrate the effect of using different interpolators in Fig.~\ref{fig1}
(taken from \cite{derivative}), where we compare Euclidean correlators for 
different interpolators with the quantum numbers of the pion. The correlators
are plotted as a function of the Euclidean time $t$, normalized such that they
are 1 at $t=1$. Since we use a logarithmic scale on the vertical axis,
the exponential factor $\exp(-t M)$ gives rise to straight lines with slope
$-M$. It is obvious, that beyond $t=5$ the different correlators all show the
same slope which corresponds to the mass $M$ of the ground state. However, for
small $t$ the correlators have a different admixture of exponential terms
$\exp(-t M^\prime)$, where $M^\prime > M$ is the mass of an excitation with the
same quantum numbers. If an interpolator couples strongly to such an
excitation, this produces a steeper slope at small $t$.

The central idea of the variational method is to use not only a single
one of the possible correlators, but a whole $r \times r$
matrix of correlators,
\begin{equation}
C_{ij}(t) \; = \; \langle O_i(t) O_j(0)^\dagger \rangle \quad , \quad
i,j = 1,\,2\, ... \, r \; ,
\end{equation}
where each of the interpolators $O_i, \, i = 1,2, \, ... \, r$ has the quantum
numbers of the channel one is interested in. It is obvious, that such a
correlation matrix extracts more information from the system than a single
correlator. 

The determination of the physical observables 
from the correlation matrix can be cast into an
elegant form \cite{variation2}. One considers the generalized eigenvalue
problem 
\begin{equation}
C(t) \, \vec{v}^{(n)} \; = \; \lambda(t)^{(n)}\ C(t_0)\ \vec{v}^{(n)} 
\; , 
\label{genevalprob}
\end{equation}
where $t_0 < t$ is a timeslice used for
normalization. Exploring the particular form of  
the spectral representation of the correlation matrix 
(compare Eq.~(\ref{specsum})),
\begin{equation}
C_{ij}(t) \; = \; 
\sum_n \langle 0 | \widehat{O}_i | n \rangle 
\langle n | \widehat{O}^\dagger_j | 0 \rangle \, e^{-t \,M_n} \; ,
\label{twopoint}
\end{equation}
one can show, that the ordered eigenvalues
$\lambda^{(1)} > \lambda^{(2)} > \lambda^{(3)} > \, ...$ behave as
\begin{equation}
\lambda^{(n)}(t) \; = \; 
e^{-(t-t_0)M_n}\ [1+\mathcal{O}(e^{-(t-t_0)\Delta_n})] \; ,
\label{evals}
\end{equation}
where $M_n$ is the mass of the $n$-th state and $\Delta_n$ the distance 
of $M_n$ to the neighboring mass. Thus the ground- and excited
states are disentangled, and each mass appears in an individual eigenvalue.
The largest eigenvalue corresponds to the ground state mass $M_1$, the second
largest eigenvalue to the mass $M_2$ of the first excitation et cetera.

The fact that in Eq.~(\ref{evals}) the ground- and excited states are
separated into individual channels allows for simple two-parameter
fits of the eigenvalues. The corresponding ground- and excited state masses
can be extracted in a stable and transparent way. 

\begin{figure}[t]
\begin{center}
\includegraphics[width=80mm,clip]{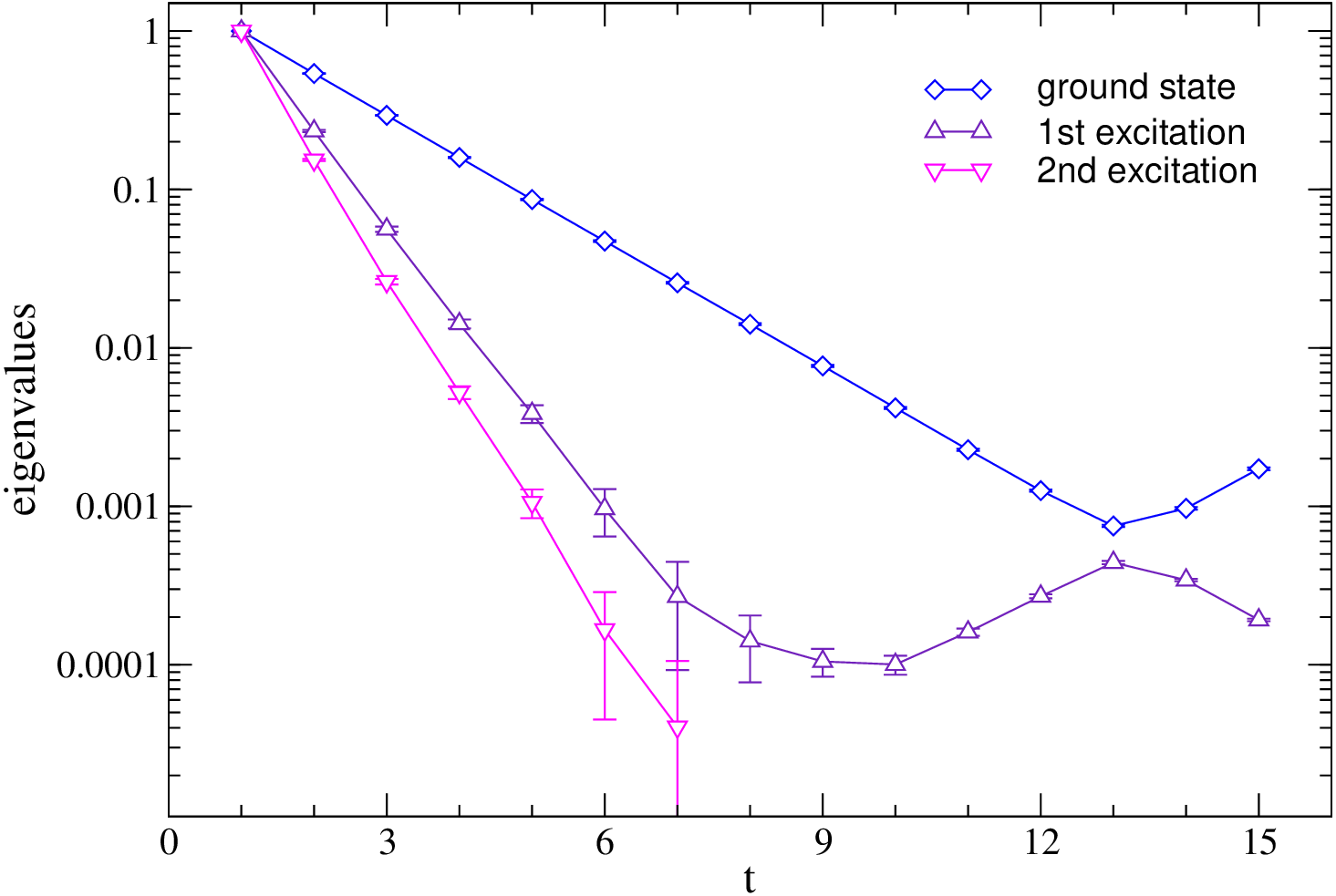}
\end{center}
\caption{Eigenvalues of the correlation matrix as a function of the Euclidean 
 time. The data are again for the example of the pion, already used in Fig.~1.}
\label{fig2}       
\end{figure}

We illustrate the power of the variational method in Fig.~\ref{fig2}, where 
we plot the eigenvalues of the correlation matrix, again showing the example of
the pion already used in Fig.~\ref{fig1}. The figure clearly demonstrates that,
when plotted on a logarithmic scale, the slopes for the eigenvalues are
different, corresponding to the different masses of the ground- and excited states.
 
\subsection{Constructing a basis of interpolators}

As any variational method, also the application to the excited states problem
can only be as good as the basis of interpolators $O_i$ one uses. Good
interpolators should have several properties: 1) They should generate states
from the vacuum that are as orthogonal to each other as possible; 2) These
states should have a large overlap with the physical states; 3) The
construction of the interpolators should be such that they can be implemented
numerically in a cost-efficient way. 

Several remarks are in order here: The true structure of the wave function of a
physical state is unknown. Thus for constructing interpolators with a ``large
overlap'' with the true physical wave function, only an educated guess, e.g., 
based on models is possible. However, even with a poor guess the method 
gives rise to the correct result, but the quality of the data might be poor. 
In practice, typically a large set of interpolators is implemented and subsequently the
set of interpolators is reduced such that only the combination with the best
signal to noise ratio is kept. If, e.g., an interpolator couples only very
weakly to physical states, it will mainly contribute noise to the data, and
should be left out in the analysis. An
important consistency check of the method is to compare the results from
different choices of the interpolators and to ensure that they agree within
error bars. 

What are the different building blocks we can use for constructing hadron
interpolators? The example of the nucleon, given in Eqs.~(\ref{chi1}),
(\ref{chi2}), illustrates that different Dirac structures may be used to obtain
different interpolators with the same quantum numbers. This freedom typically
gives rise to two or three (depending on the channel) different hadron 
interpolators with the same quantum numbers $I,J,P$ and $C$ (where
applicable). The different Dirac structures are an important
ingredient, but obviously not enough for analyzing the tower of all
excitations. It is generally believed, that for a proper description of excited
hadrons non-trivial spatial wave functions are essential. 

Recently the lattice community started to systematically explore the
possibility of implementing non-trivial spatial wave functions
\cite{wavefunct1}-\cite{derivative} 
in hadron interpolators. Two different approaches for constructing such
wavefunctions on the lattice are used: 

Basak {\it et al} 
have constructed large sets of possible hadron interpolators
by using quark fields on lattice sites displaced relative to each other,
typically involving nearest or next-to-nearest neighbors
\cite{wavefunct2}. The operators were
classified with respect to irreducible representations of the symmetry group
of the hyper-cubic lattice, and in in this way the quantum numbers were
assigned. 

A different strategy was followed in the work of the Graz-Regensburg group
\cite{wavefunct1}-\cite{baryons}, \cite{derivative}. This approach is based 
on so-called Jacobi smeared quark sources \cite{jacobi}, which is essentially
a gauge covariant diffusion process, that leads to an s-wave type of wave
function for an individual quark. Combining Jacobi-smeared sources of
different width allows for radial wave functions with nodes. Application of
an additional covariant derivative gives rise to p-wave type wave functions
\cite{derivative}. 

\begin{figure}[t]
\begin{center}
\includegraphics[width=70mm,clip]{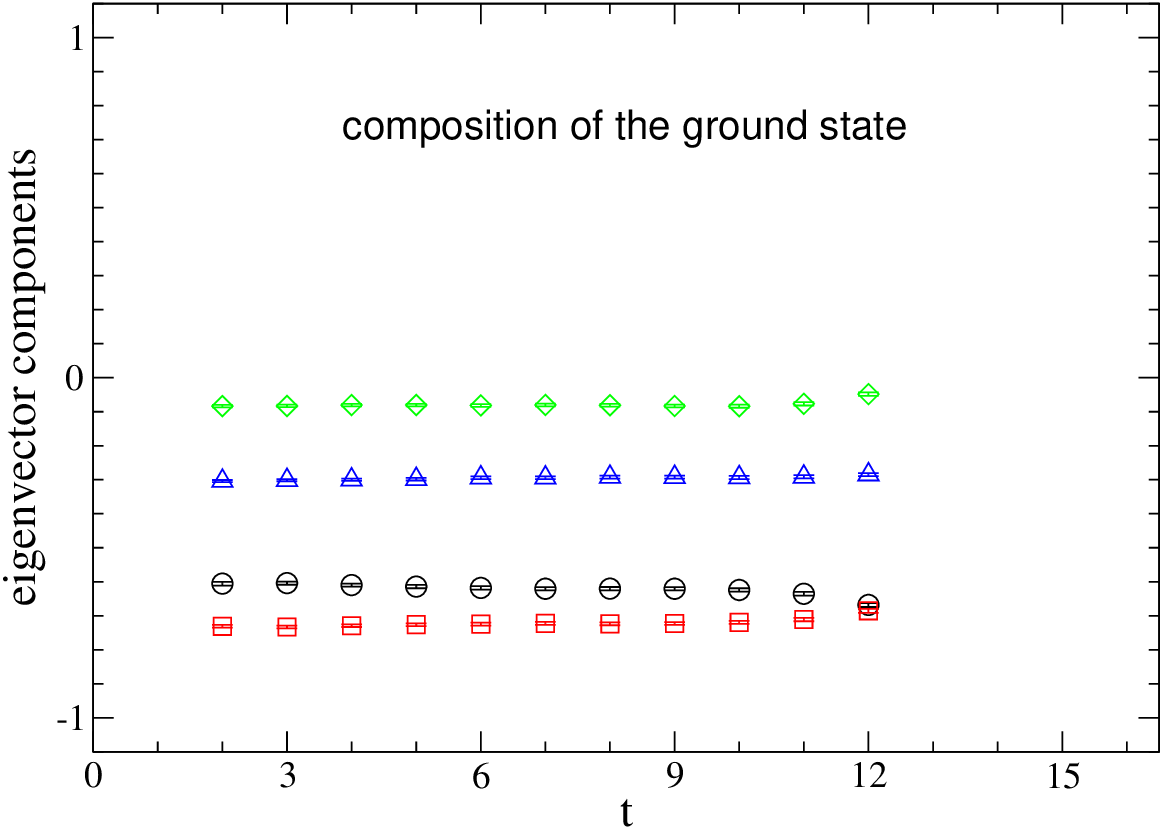}
\end{center}
\caption{Entries of the first eigenvector of the generalized eigenvalue
  problem for the pion example already used in Figs.~1,2. 
The entries are plotted as a function of the
  Euclidean time $t$.}
\label{vectorplot}       
\end{figure}

It is important to understand that the different interpolators one uses in
the variational approach are just the basis one offers to the system. The
relative weights of the basis elements are not prescribed, 
but come out of the variational
calculation. Once the generalized eigenvalue problem 
(\ref{genevalprob}) is solved
and the eigenvectors $\vec{v}^{(n)},\, n = 1,2 \, ... \, r$ are known, one can
define new interpolators $\widetilde{O}^{(n)}$ as linear combinations 
of the original interpolators $O_i$,
\begin{equation}
\widetilde{O}^{(n)} \; = \; \sum_{i=1}^r \, {v^{(n)}_i}^* \, O_i \; ,
\end{equation}
with the complex conjugate (denoted by $^*$) entries ${v^{(n)}_i}^*$ 
of the $n$-th
vector as coefficients. The new interpolators $\widetilde{O}^{(n)}$ are 
optimal within the given basis of interpolators $O_i$ in the sense that they
give rise to orthogonal correlation functions, 
\begin{equation}
\langle \widetilde{O}^{(m)}(t) \, \widetilde{O}^{(n) \, \dagger} (0) \rangle 
\; = \; \lambda^{n}(t) \, \delta_{n,m} \; ,
\end{equation}
as can be shown in a few lines of algebra. Thus the variational method
determines through the eigenvectors which linear combinations of the basis
interpolators best describe a physical state. In principle the eigenvectors can
be functions of the Euclidean time $t$. It was observed, that the 
eigenvectors are essentially independent of $t$ in the range where the
eigenvalues are dominated by a single exponential. In Fig.~\ref{vectorplot} we
illustrate this behavior for the eigenvector corresponding to the ground
state, again using the pion example. 
Comparison with Fig.~2, where the corresponding eigenvalues are
plotted, shows that indeed the ground state eigenvalue falls on a straight
line up to $t=12$, exactly the range where the eigenvector shown in Fig.~3
displays plateaus for its entries. The property that also the eigenvectors
show a signal stable in $t$ is an important technical tool for identifying the
different states and for the determination of fit ranges.    

\section{Results and future challenges}

\subsection{Some selected results}

Having outlined the ingredients of an excited state calculation in lattice
QCD, let us now present some selected results. We begin with discussing the
outcome of a quenched calculation for excited mesons \cite{mesons}. The
calculation is based on meson interpolators constructed with the Jacobi 
smearing techniques outlined in the last section. The quark propagators were
computed for several different values of the light quark masses and the
results for the meson spectra were extrapolated to the chiral limit. The scale
was set with the Sommer parameter and the value of the strange quark mass was
determined from the $K^+$ meson. 

\begin{figure}[t]
\begin{center}
\includegraphics[width=70mm,clip]{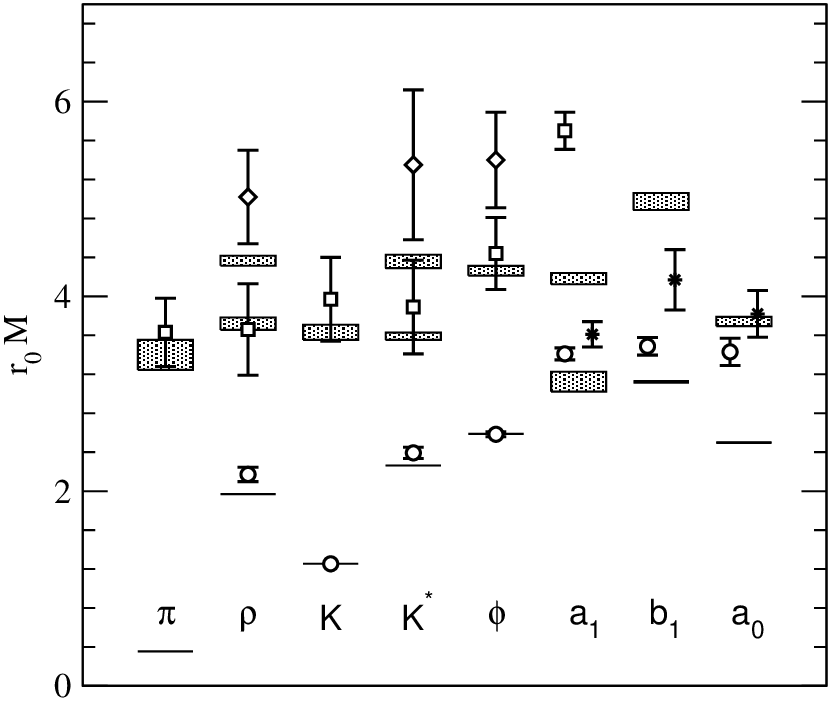}
\end{center}
\caption{Final results for excited mesons taken from \cite{mesons}. The
  horizontal bars are the experimental data, the symbols correspond to the
  lattice results. All masses are given in units of the Sommer scale 
$r_0 = 0.5fm$.}
\label{figmeson}       
\end{figure}

In Fig.~\ref{figmeson} we show the results for the mass spectrum in different 
meson channels. The bars are the experimental numbers and the symbols are the
lattice results. The errors for the lattice data are statistical errors.
The plot shows that for pseudoscalar and vector mesons the lattice
results agree well with the experimental values. The ground state masses
typically are correct within 5 percent (10\% for the $\rho$-meson) and one or
two excited states can be calculated which for the first excitations 
agree with the experimental values within error bars. For the
axialvector/tensor mesons the results are less convincing and we believe that
here the set of basis interpolators is still not rich enough. Finally for the 
scalar channel the lattice data seem to coincide with the experimental mass of
the first excitation. Understanding the true nature of the scalar meson ground
state is an interesting story of its own -- it could, e.g., be a tetraquark
state -- and for a snapshot of the ongoing lattice work on this problem we
refer the reader to the recent review \cite{lattreview}.  

Let us now come to discussing some results for baryons \cite{baryons}. This
quenched calculation is again based on the variational method with basis 
interpolators
constructed with Jacobi smearing. As for the meson case, the light
quark masses were extrapolated to the chiral limit and the strange quark mass
was set using the $K^+$ as input. 

\begin{figure*}[t]
\begin{center}
\includegraphics[width=80mm,clip]{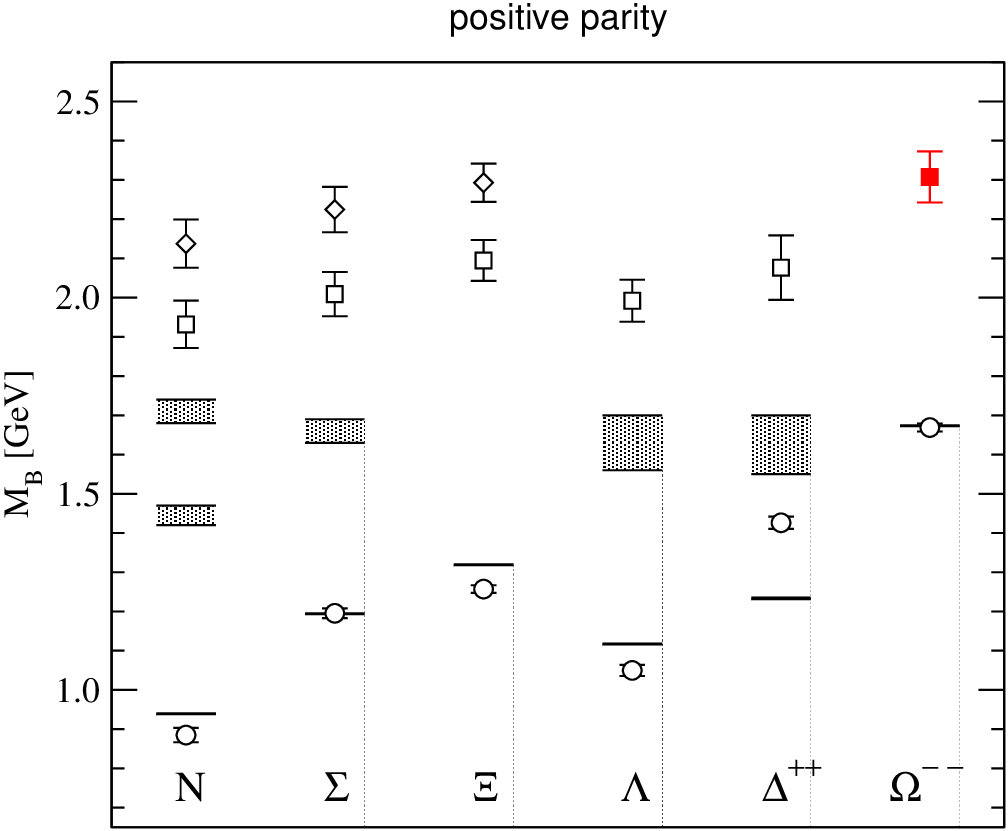}
\hfill
\includegraphics[width=80mm,clip]{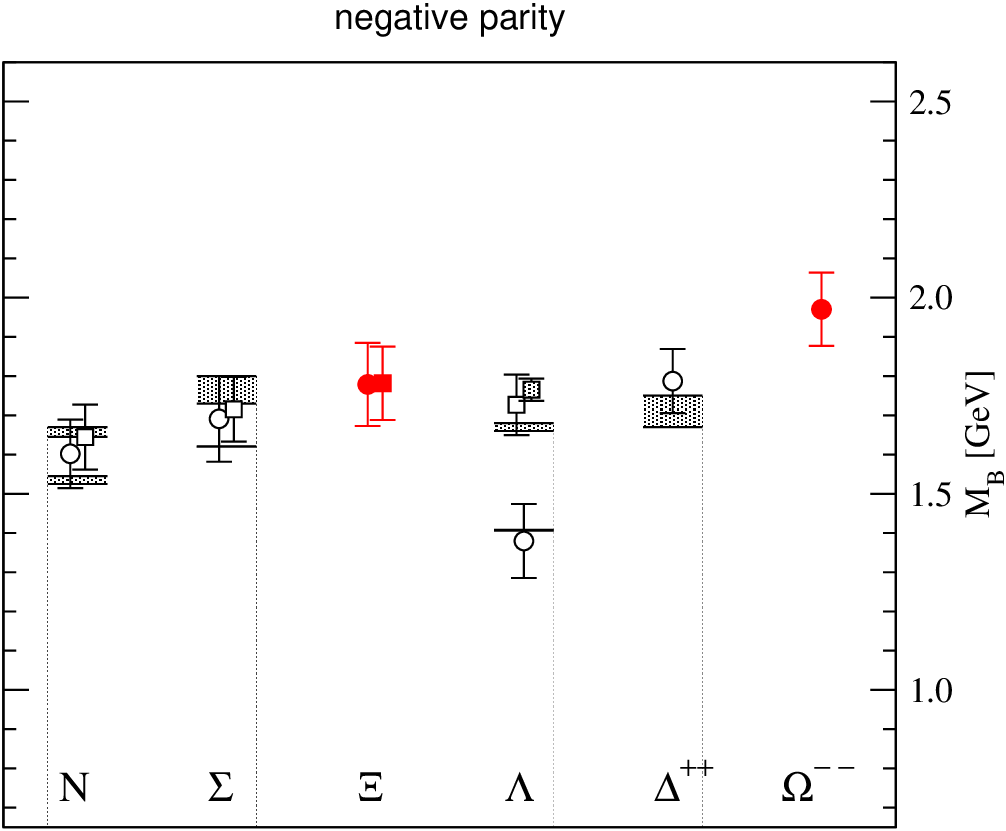}
\end{center}
\caption{Final results for excited baryons taken from \cite{baryons}. The
  horizontal bars are the experimental data, the symbols correspond to the
  lattice results. In the lhs.~plot we show the results for positive parity
  baryons, while the rhs.~is for negative parity.}
\label{figbaryon}       
\end{figure*}

Fig.~\ref{figbaryon} shows the results for positive parity in the lhs. plot,
while the rhs.~is for negative parity. For both parities the ground state
masses come out reasonably well - typically with less than 5 percent deviation
from the experimental numbers. As for the meson case one is wondering at this
point why the quenched approximation is working so well. An exception is the
positive parity $\Delta^{++}$ ground state which typically comes out about 
20\% too high \cite{earlyexcite}. A completely different story are the
excited positive parity states. They come out typically 25\% too high. 
Several possible reasons for this failure to describe these Roper-like states
have been put forward, such as the quenched approximation, too small
spatial lattice volumes and the fact that the light quark masses used still 
might be to large to capture the chiral dynamics necessary to get the masses
right \cite{earlyexcite}. In general the opinion is, that the Roper 
puzzle will be solved only when dynamical simulations with light quarks on
large volumes are completed. This is certainly an expensive enterprise, but the
whole lattice community is pushing in this direction, and once these data
are available, they will certainly be used for studying excited hadrons. 

To conclude the discussion of the baryons with a more positive aspect, 
we stress that the negative parity states come out very nicely. The
lattice typically produces results that are on top of the experimental
numbers. For a better resolution of the splitting between the ground- and the first
excited state the error bars will have to be reduced. Nevertheless the results
are sufficiently reliable, such 
that the lattice can make a prediction for states in the $\Xi$ and
$\Omega^{--}$ channels, where the corresponding masses are not yet well 
established experimentally (filled symbols in Fig.~\ref{figbaryon}).

\subsection{Upcoming challenges}   

Let us finally come to the challenges that are waiting for excited state
calculations on the lattice in the near future. We have already stressed the
point that the quenched approximation is the big unknown in the current
results for excited hadrons. We have also addressed the problem with the
mixing to scattering states which has to be faced when light dynamical
fermions are considered. In principle the techniques for disentangling 
scattering and bound states are ready to go \cite{scattertechniques}: For
scattering states the relative momentum is quantized in multiples of the 
Matsubara frequency $2\pi/L$, where $L$ is the spatial extent of the
lattice. Thus the energies of scattering states show a power-like dependence
on $1/L$ which is absent for bound states, and comparing the results on
different volumes allows to identify scattering states. However, this is an
expensive strategy which so far has been tested mainly in low-dimensional
models and scalar field theories \cite{variation2,scatter}. 

Another interesting challenge is the evaluation of matrix elements for excited
states. Here some conceptual problems still have to be addressed, but in principle
it should be possible to formulate this problem within the variational
approach. A first appetizer of what the lattice could achieve in this
direction is the calculation of the decay constant for $\pi(1300)$ presented
in \cite{pidecay}. 

\vskip5mm
\noindent
{\bf Acknowledgements:} I would like to thank  
D.~Br\"ommel, T.~Burch, J.~Danzer, L.~Glozman, C.~Hagen, D.~Hierl, M.~Limmer, 
C.B.~Lang, D.~Mohler, S.~Prelovsek,  S.~Schaefer and A.~Sch\"afer,
who I have collaborated with on excited hadron phyics over the last few
years. Some of the results presented here were obtained on the parallel
computers at the Leibniz Rechenzentrum, Garching, and at the ZID,
Universit\"at Graz.

\end{document}